\documentclass[iop]{emulateapj}
\usepackage{apjfonts}
\usepackage[breaklinks,colorlinks,urlcolor=black,citecolor=black,linkcolor=black]{hyperref}

\newcommand{\etal}{et~al.\/}

\newcommand{\sqdegs}{square degrees}

\newcommand{\mjypbm}{{\rm mJy~beam}\ensuremath{^{-1}}}

\newcommand{\jypbm}{{\rm Jy~beam}\ensuremath{^{-1}}}

\newcommand{\eg}{e.\,g.\,}

\newcommand{\esh}{erg s\ensuremath{^{-1}} Hz\ensuremath{^{-1}}}

\newcommand{\rchisq}{\ensuremath{\chi^2_\nu}}

\newcommand{\fracmod}{\ensuremath{\sigma_{\rm S} / \bar{\rm S}}}

\newcommand{\antth}{ANT\,131121A}
\newcommand{\antfo}{ANT\,140323A}
\newcommand{\sigback}{\ensuremath{\sigma_{\rm sky}}}

\shorttitle{MWA Follow-up of ANTARES Triggers}
\shortauthors{Croft \etal}

\begin{document}
\title{Murchison Widefield Array Limits on Radio Emission from ANTARES Neutrino Events}
\author{S.~Croft\altaffilmark{1,2}, 
D.~L.~Kaplan\altaffilmark{3}, 
S.~J.~Tingay\altaffilmark{4,5}, 
T.~Murphy\altaffilmark{5,6}, 
M.~E.~Bell\altaffilmark{7}, 
A.~Rowlinson\altaffilmark{5,8,9},
for the MWA Collaboration\\ \vspace{0.2cm} 
S.~Adri\'an-Mart\'inez\altaffilmark{10}, 
M.~Ageron\altaffilmark{11}, 
A.~Albert\altaffilmark{12}, 
M.~Andr\'e\altaffilmark{13}
% ...
for the ROTSE Collaboration
}

\altaffiltext{1}{University of California, Berkeley, Astronomy Dept., 501 Campbell Hall \#3411, Berkeley, CA 94720, USA }
\altaffiltext{2}{Eureka Scientific, Inc., 2452 Delmer Street Suite 100, Oakland, CA 94602, USA}
\altaffiltext{3}{Department of Physics, University of Wisconsin-Milwaukee, 1900 E. Kenwood Boulevard, Milwaukee, WI 53211, USA}
\altaffiltext{4}{International Centre for Radio Astronomy Research, Curtin University, Bentley, WA 6102, Australia}
\altaffiltext{5}{ARC Centre of Excellence for All-sky Astrophysics (CAASTRO)}
\altaffiltext{6}{Sydney Institute for Astronomy, School of Physics, The University of Sydney, NSW 2006, Australia}
\altaffiltext{7}{CSIRO Australia Telescope National Facility, PO Box 76, Epping, NSW 1710, Australia}
\altaffiltext{8}{Anton Pannekoek Institute for Astronomy,  University of Amsterdam, Science Park 904, 1098 XH Amsterdam, The Netherlands}
\altaffiltext{9}{ASTRON, The Netherlands Institute for Radio Astronomy, Postbus 2, 7990 AA, Dwingeloo, The Netherlands}
\altaffiltext{10}{Institut d'Investigaci\'o per a la Gesti\'o Integrada de les Zones Costaneres (IGIC) - Universitat Polit\`ecnica de Val\`encia. C/  Paranimf 1, 46730 Gandia, Spain}
\altaffiltext{11}{Aix Marseille Universit\'e, CNRS/IN2P3, CPPM UMR 7346, 13288, Marseille, France}
\altaffiltext{12}{GRPHE - Universit\'e de Haute Alsace - Institut universitaire de technologie de Colmar, 34 rue du Grillenbreit BP 50568 - 68008 Colmar, France}
\altaffiltext{13}{Technical University of Catalonia, Laboratory of Applied Bioacoustics, Rambla Exposici\'o, 08800 Vilanova i la Geltr\'u, Barcelona, Spain}
% ...

\tabletypesize{\scriptsize}

\begin{abstract}
We present a search, using the Murchison Widefield Array (MWA), for electromagnetic counterparts to two candidate high energy neutrino events detected by the ANTARES neutrino telescope in 2013 November and 2014 March. These events were selected by ANTARES because they are consistent, within 0\fdg4, with the locations of galaxies within 20\,Mpc of Earth. Using MWA archival data at frequencies between 118 and 182\,MHz, taken $\sim 20$~days prior to, at the same time as, and up to a year after the neutrino triggers, we look for transient or strongly variable radio sources consistent with the neutrino positions. No such counterparts are detected, and we set a $5 \sigma$ upper limit for low-frequency radio emission of $\sim 10^{37}$\,erg s$^{-1}$ for progenitors at 20\,Mpc. If the neutrino sources are instead not in nearby galaxies, but originate in binary neutron star coalescences, our limits place the progenitors at $z \gtrsim 0.2$. While it is possible, due to the high background from atmospheric neutrinos, that neither event is astrophysical, the MWA observations are nevertheless among the first to follow up neutrino candidates in the radio, and illustrate the promise of wide-field instruments like MWA to detect electromagnetic counterparts to such events. 
\end{abstract}


\section{Introduction}

Neutrinos are believed to be emitted by a range of astrophysical sources \citep{chiarusi:10,anchordoqui:10}, including transient sources such as gamma-ray bursts (GRBs), core-collapse supernovae (CCSNe), active galactic nuclei (AGNs), or microquasars. Neutrinos provide a powerful probe of high-energy astrophysical environments, because they are unaffected by magnetic fields, and are extremely unlikely to be absorbed by material between the source and observer. These same properties make them very challenging to detect, even by the largest of the current generation of neutrino observatories, and contaminating background signals are high. However, if their directions can be localized, they have the potential to point directly back to the astrophysical accelerators in which they are created. 

Even so, typical positional uncertainties from neutrino telescopes are large enough to encompass many potential electromagnetic (EM) counterparts. A solution to dramatically decrease association ambiguity is to search for transient EM emission that is spatially and temporally consistent with neutrino events. However, aside from neutrinos from the Sun, the only astronomical source so far associated with a neutrino detection (in the tens of MeV energy range) is SN\,1987A \citep{sn1987a,bionta:87,hirata:87,alexeyev:88} --- although recently \citet{kadler} reported a blazar outburst coincident with a PeV-energy neutrino event. Timely multi-wavelength follow-up of neutrino candidates is key in order to attempt to identify the progenitors of astrophysical neutrinos.

The two most sensitive neutrino telescopes currently operating are ANTARES \citep{antares} and IceCube \citep{icecube}. Both search for Cherenkov radiation from secondary particles produced from cosmic neutrinos with energies $> 100$\,GeV.
For IceCube \citep{icecubehe}, located at the South Pole, neutrinos from the southern sky are observed as downward-going. Below a PeV, these neutrinos are selected with a vetoing technique that favors the detection of showering events, for which the detector has an angular resolution of only $10\degr - 15\degr$.

ANTARES, located 40\,km off the southern coast of France in the Mediterranean Sea, views the southern sky via upward-going neutrino-induced muon tracks, with a characteristic resolution (50\%\ error circle) of 0\fdg4 \citep{adrianmartinez:14}. The detector produces the best limits on neutrino emissions for point-like objects in most of this southern sky region, and hence EM follow-up efforts are concentrated there. A dedicated alert system, TAToO \citep{tatoo}, is triggered when a candidate special neutrino event is detected: a single high-energy neutrino; a neutrino in the direction of a local galaxy; or at least two neutrinos coincident in space and time \citep{antaresearly}.

In this analysis, we searched the MWA archives for observations that were coincident in time and position with neutrino triggers from ANTARES from mid 2013 to mid 2015. Two events, \antth\ and \antfo, were found to occur within the MWA field of view (Table~\ref{tab:triggers}), when the MWA happened to be observing the trigger position just prior to, during, and immediately after the trigger time. 

A brief description of the two events and optical follow-up is given in Section~\ref{sec:neutrino_events}. The MWA follow up is presented in Section~\ref{sec:MWA_ana}. Limits on progenitors, as well as prospects for future work, are presented in Sections~\ref{sec:limits} and \ref{sec:future}, respectively.

\section{ANTARES neutrino events and optical follow-up}\label{sec:neutrino_events}

ANTARES detects $2 - 3$ neutrino candidates per day on average. From mid 2013 to mid 2015, more than 60 ANTARES events satisfied one of the three special triggers discussed above, and a TAToO alert was issued. For many of these alerts, a network of robotic optical telescopes started observations as soon as possible (prompt strategy) and continued up to two months (long-term strategy) after the neutrino detection. These strategies are well-suited to the search for rapidly-varying transient sources, such as GRB afterglows, and slowly-varying sources, such as CCSNe. 

Both triggers with simultaneous MWA observations were among the $\sim 30$ selected with the ANTARES directional trigger between mid 2013 and mid 2015. Such triggers have directions consistent ($< 0\fdg4$) with the positions \citep{gwgc} of galaxies within 20\,Mpc of Earth. Two galaxies match in each case: NGC\,1374 and ESO\,358-015 match \antth, and ESO\,499-037 and PGC\,29194 match \antfo. PGC\,29194 (the Antlia Dwarf Galaxy), at a distance of 1.3\,Mpc, is located just 6\arcmin\ from the neutrino position.

Both neutrino events also had optical follow-up. For \antth, 12 observations of 6 images were performed with the 0.25-m TAROT telescope in Chile from $2 - 61$ days after the trigger. Optical images were analyzed with an image-subtraction pipeline \citep{antaresearly}. No transient was identified, to a limiting magnitude of $\sim 19$ \citep{antaresmid}.
For \antfo, a total of 8 images were taken with ROTSE 3b in Texas (starting $\sim$ 15\,hr after the trigger) according to the prompt strategy, and 10 images with TAROT Chile up to 45 days after the trigger according to the long-term strategy. No transient counterpart was found \citep{antaresmid,antaresearly}, to limiting magnitudes of 16.4 (prompt) and 18.7 (long-term).

\begin{deluxetable}{lllrrl}
\tablewidth{0pt}
\tabletypesize{\scriptsize}
\tablecaption{\label{tab:triggers} Details of the two ANTARES events with simultaneous MWA observations}
\tablehead {
\colhead{Trigger ID} &
\colhead{UT date} &
\colhead{UT time} &
\colhead{RA} &
\colhead{Dec} &
\colhead{Energy}\\
\colhead{} &
\colhead{} &
\colhead{} &
\colhead{(deg)} &
\colhead{(deg)} &
\colhead{(TeV)}
}
\startdata
\antth & 2013 Nov 21 & 14:58:28 & 53.5  & $-35.1$ & $\sim 1$ \\
\antfo & 2014 Mar 23 & 15:31:01 & 150.9 & $-27.4$ & $\sim 4$
\enddata
\end{deluxetable}

\section{MWA Follow-up of ANTARES Events}\label{sec:MWA_ana}

The Murchison Widefield Array (MWA), situated in Western Australia, is the Square Kilometre Array precursor at low ($80 - 300$\,MHz) radio frequencies \citep{mwadesign,mwaska}. The MWA is often used to undertake surveys, for a range of science goals including dedicated \citep[\eg,][]{bmk+13,murphy:15} and commensal \citep[\eg][]{mwafrb,tingay:frb} transient searches, but it has also been used for triggered follow-up of transients at other wavelengths \citep[\eg,][]{grb150424a}. Its huge field of view (700~\sqdegs\ at 150\,MHz) also means that archival observations have a much larger chance, compared to most other radio telescopes, of serendipitously covering an event of interest. This capability is particularly valuable for follow-up of neutrino or gravitational wave \citep{singer:15} candidates, which have rather large position uncertainties.

We obtained MWA archival data for both ANTARES triggers, from periods before (Section~\ref{sec:pre}) and at the time of (Section~\ref{sec:prompt}) the trigger, in a search for prompt emission. We also searched for data over a longer range of time to look for late-time emission (Section~\ref{sec:post}).

Flagged CASA \citep{casa} measurement sets were produced using the MWA preprocessing pipeline COTTER \citep{cotter}. These were then processed by our custom imaging pipeline, which used WSCLEAN \citep{wsclean} with 40,000 CLEAN iterations to produce XX and YY polarization images with $3072 \times 3072$ $0\farcm9$ pixels. The images were amplitude and phase self-calibrated, and primary beam corrected to produce Stokes I images which formed the basis for our analysis. Catalogs were generated using Aegean \citep{aegean} and cross-matched across snapshots.

\subsection{Search for Prompt Emission}\label{sec:prompt}

For each of the two triggers, we retrieved 34 MWA datasets, in addition to observations of nearby bright radio calibrators (Pic\,A for \antth\ and Hyd\,~A for \antfo). Exposure times were 112\,s, and snapshots were taken approximately every 2\,min, from $\sim 10$\,min before the neutrino trigger to 1\,hr after (sufficiently long to probe dispersion measures $> 10^4$ pc\,cm$^{-3}$). For \antth, the central frequency for each observation was 154.255\,MHz, and the bandwidth was 30.72\,MHz, divided into 768 channels of 40\,kHz. \antfo\ had the same bandwidth and channels, but the central frequency was 182.415\,MHz. The MWA synthesized beam is $\sim 2\arcmin \times 2\arcmin$ at 154\,MHz.

One of the 34 snapshots for \antth\ failed to image adequately and was discarded. Comparison by eye of the remaining snapshot images for each trigger showed no obvious transients to be present in or near the ANTARES 90\%\ error circles, which are 1\degr\ in radius \citep{adrianmartinez:14}. Additionally, no transients (sources $\geq 5$ times brighter than the background fluctuations) were present in catalogs corresponding to a single snapshot within the ANTARES error circles.

\begin{figure*}[htp]
\centering
\includegraphics[width=0.3\linewidth,draft=false]{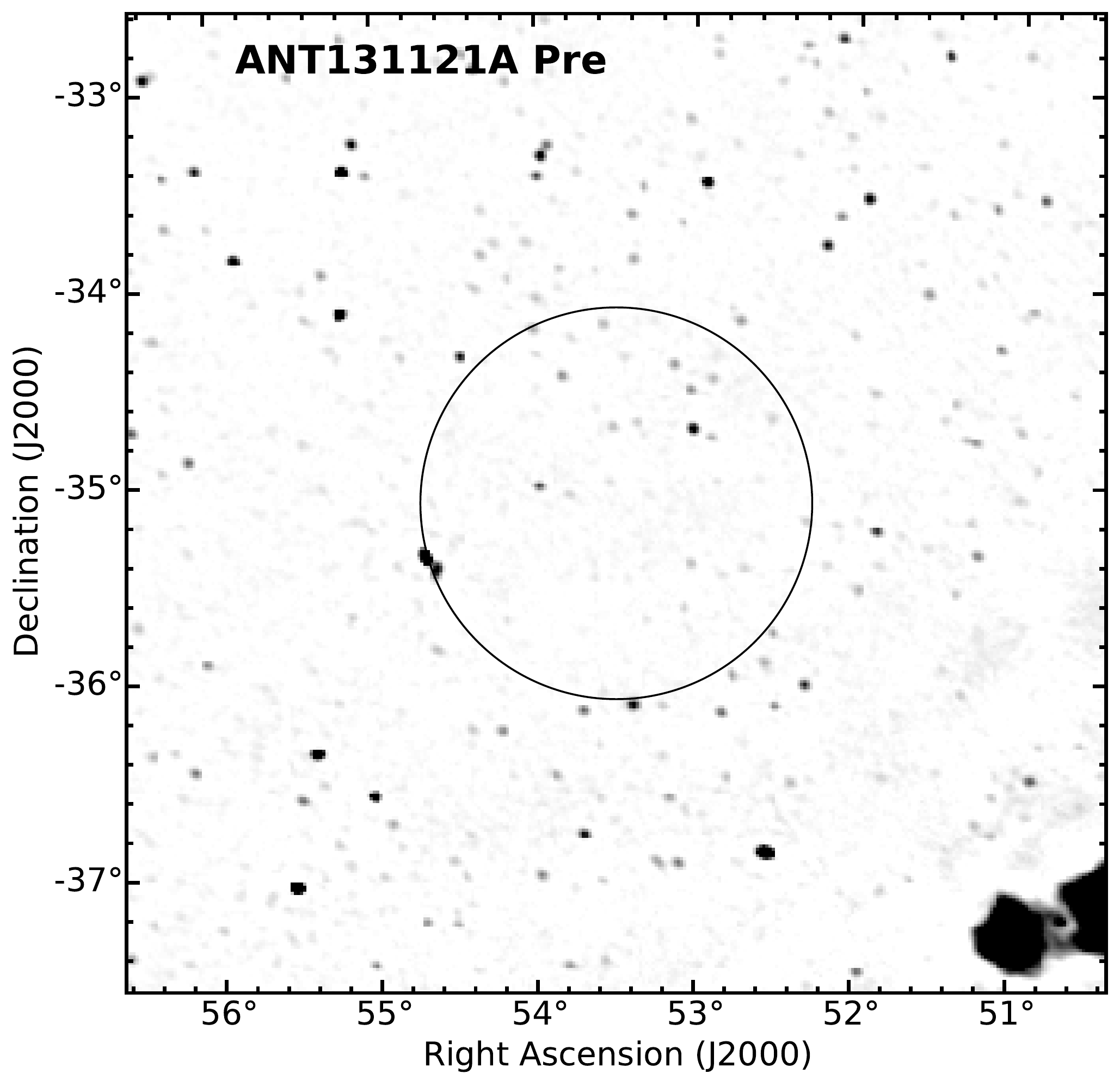}\quad%
\includegraphics[width=0.3\linewidth,draft=false]{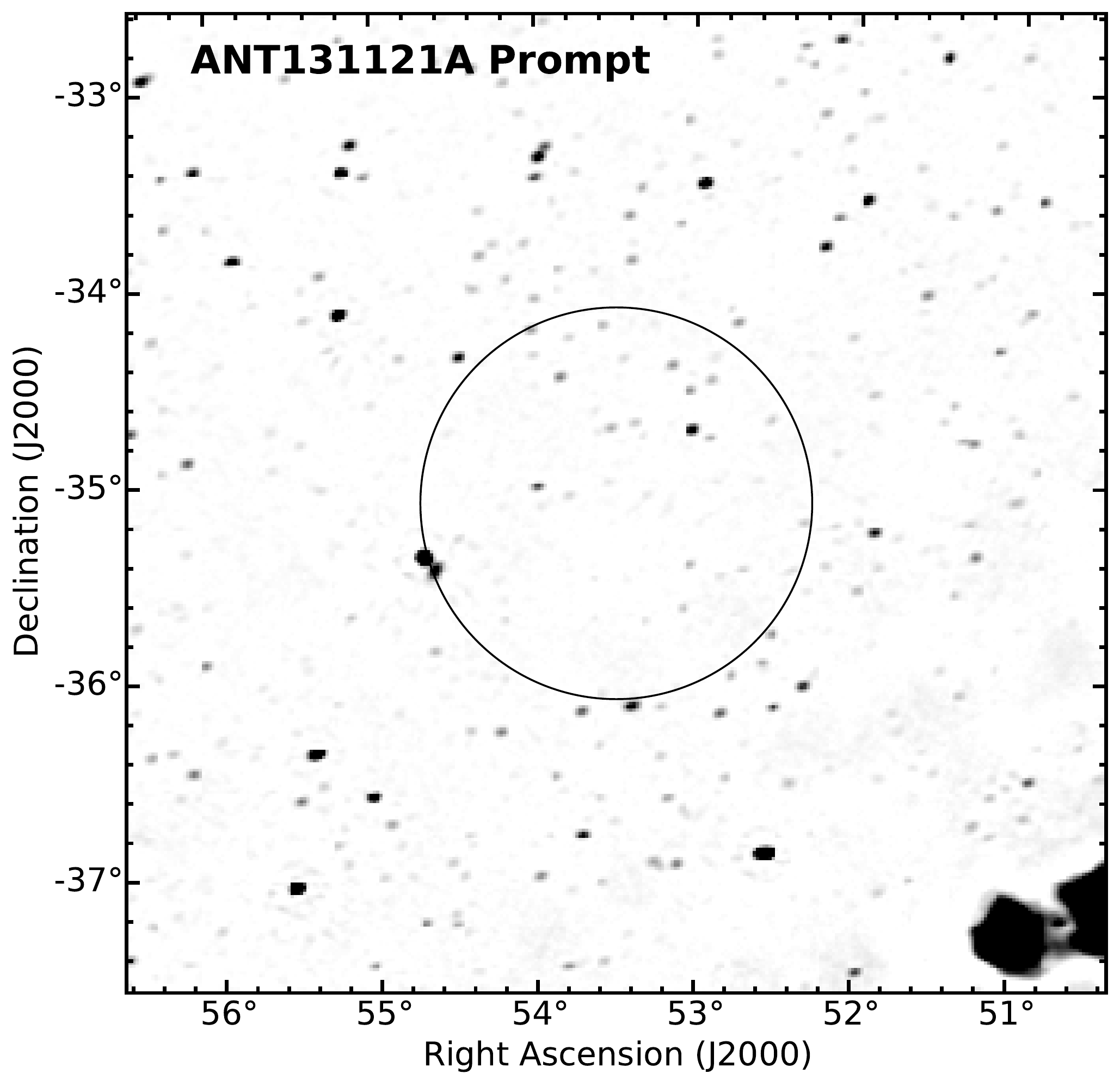}\quad%
\includegraphics[width=0.3\linewidth,draft=false]{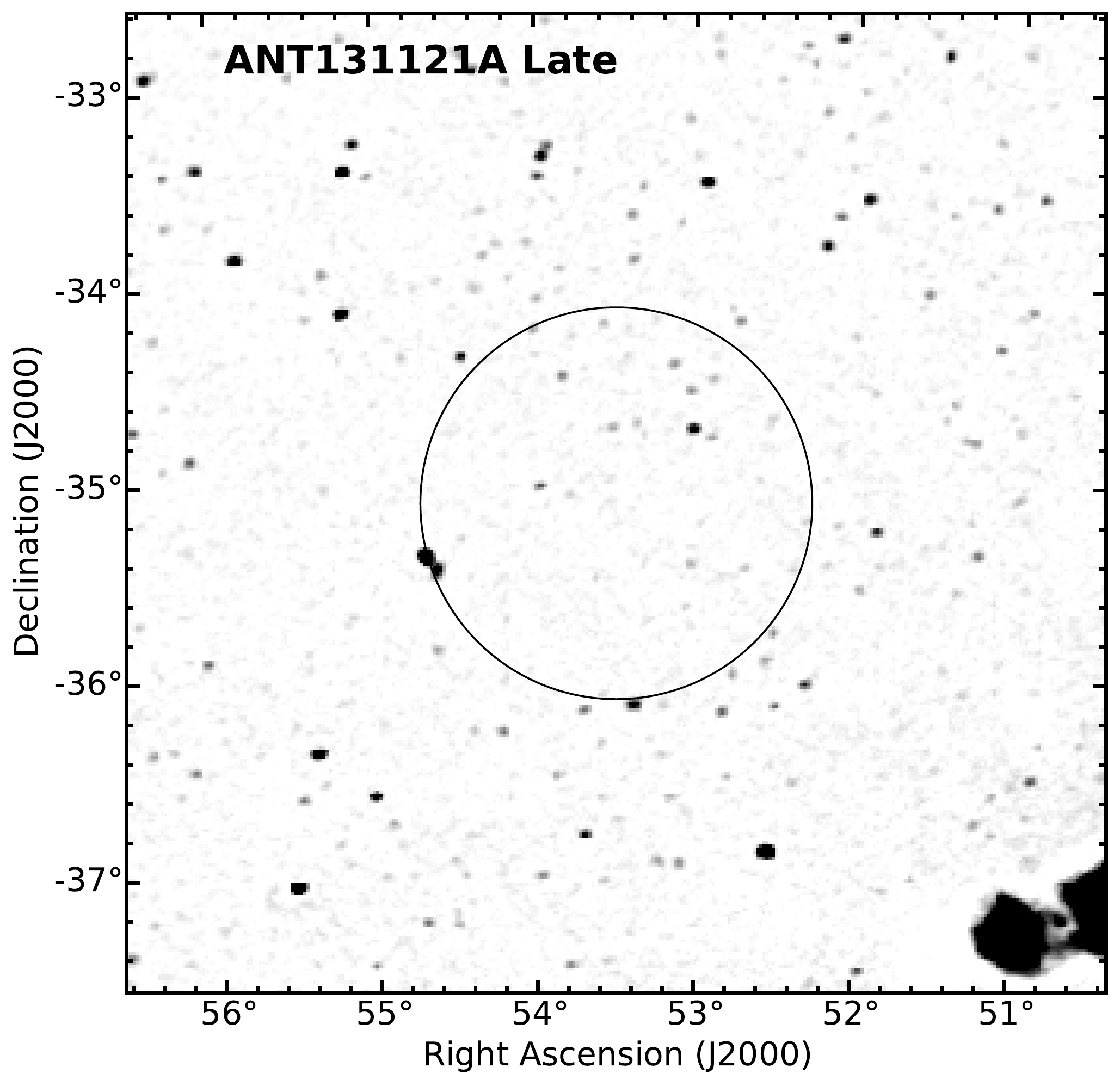}\quad%
\includegraphics[width=0.3\linewidth,draft=false]{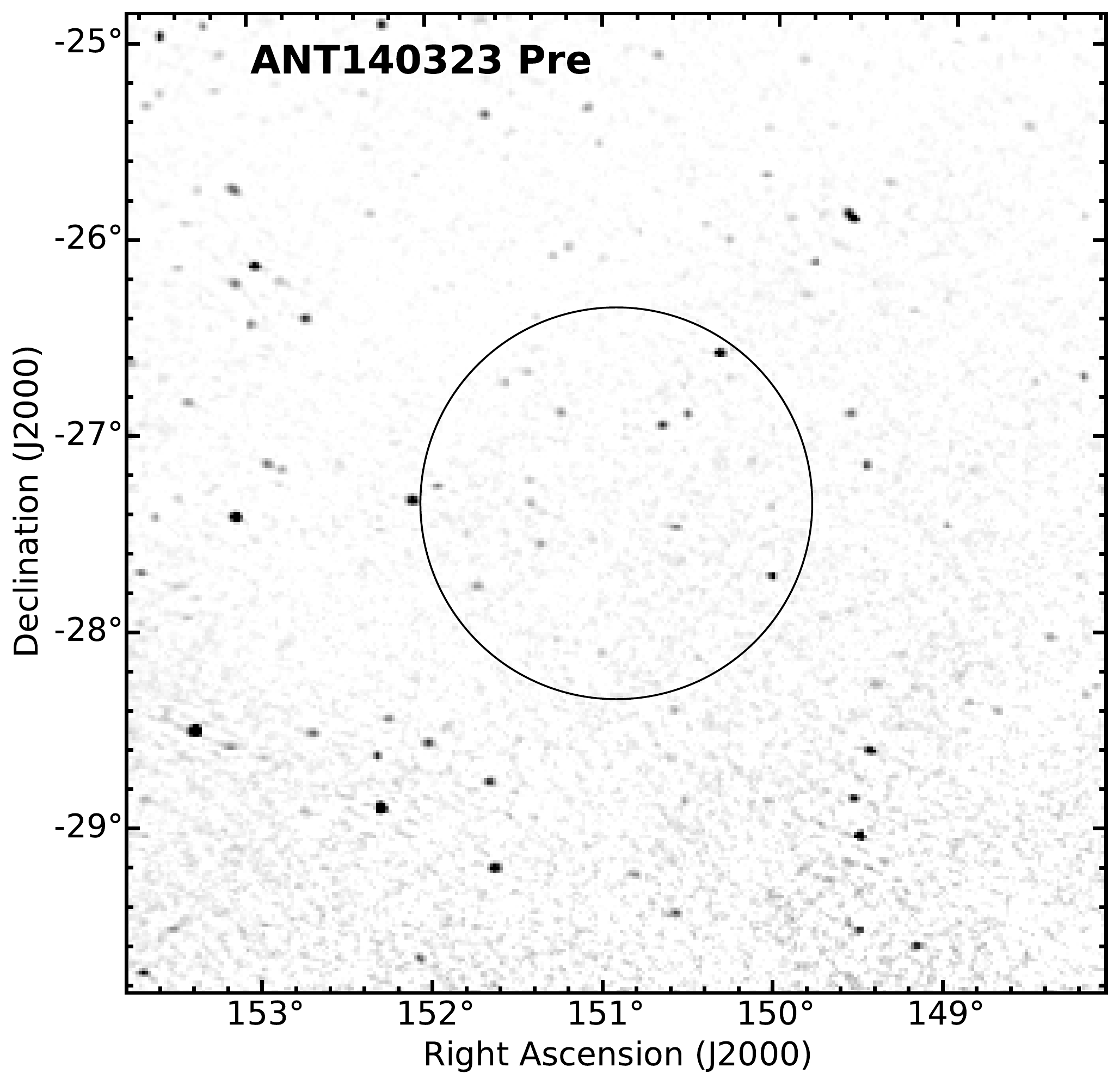}\quad%
\includegraphics[width=0.3\linewidth,draft=false]{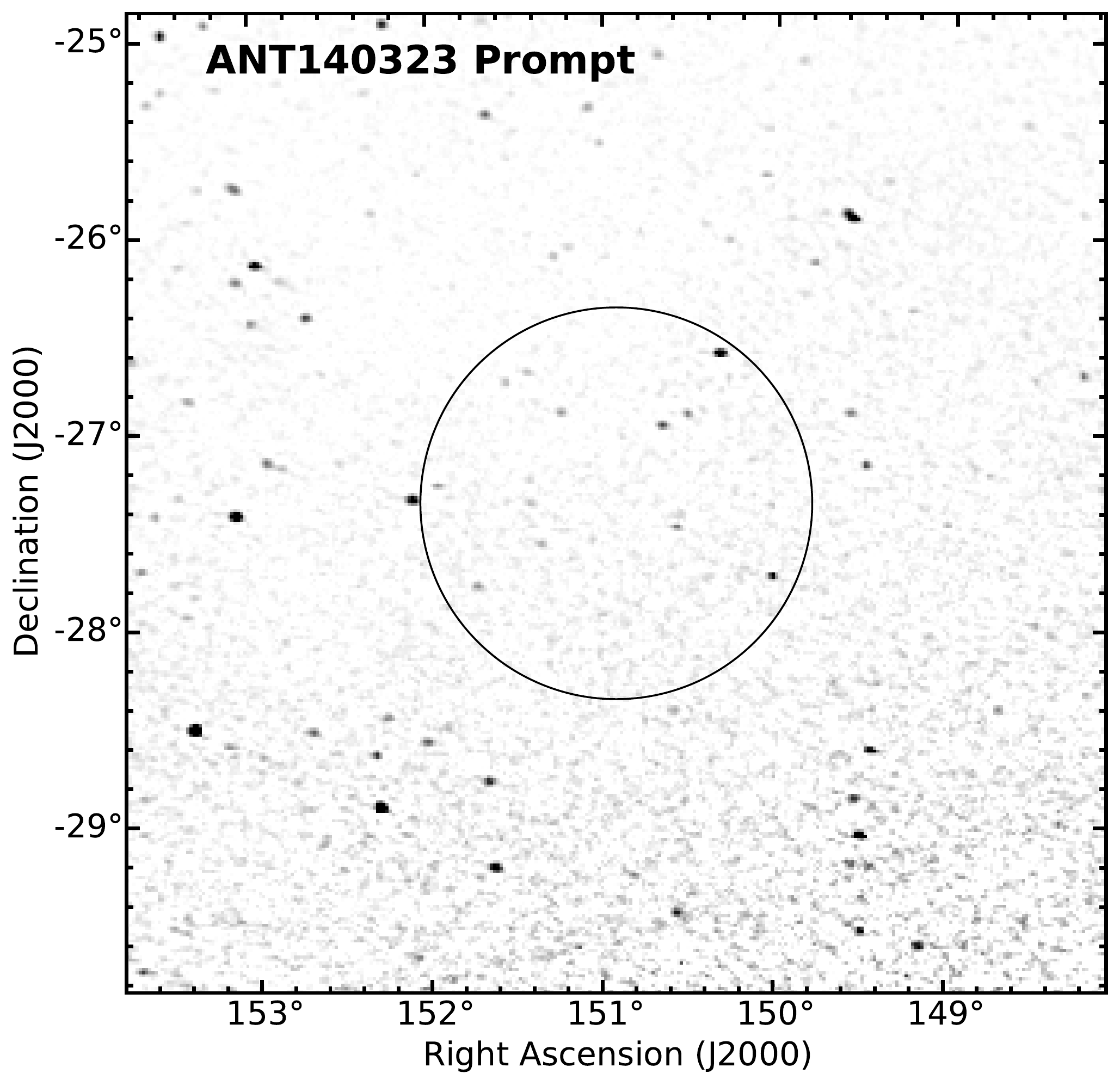}\quad%
\includegraphics[width=0.3\linewidth,draft=false]{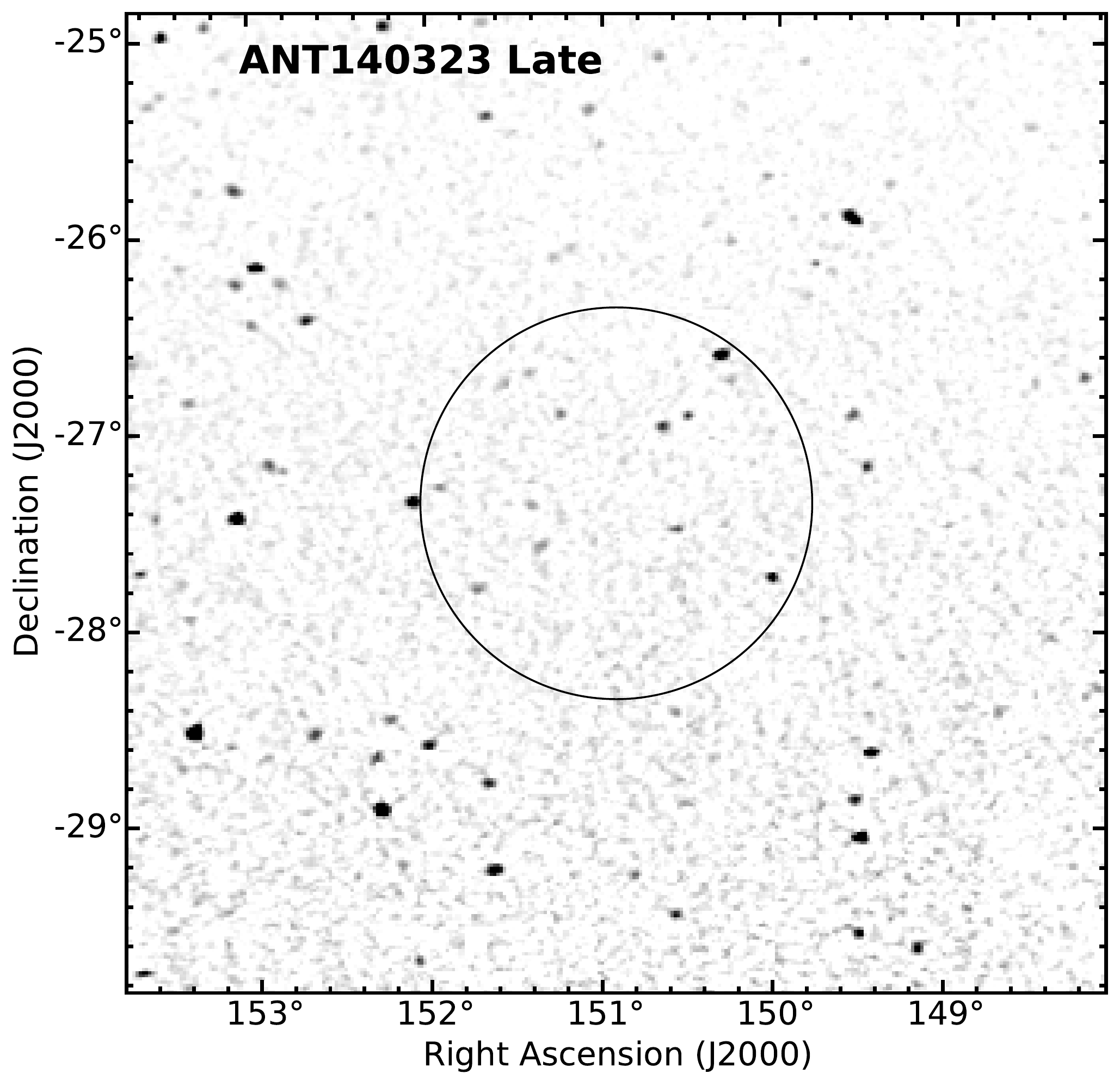}\quad%
\caption{\label{fig:deep}
$5\degr \times 5\degr$ cutouts from median-combined deep images on each of the two triggers (top: \antth; bottom: \antfo). From left to right, images were taken $\sim 20$ days prior to, at the time of, and over the course of $\sim 1$\,yr after the trigger time (Table~\ref{tab:rms}).
Some faint image artifacts are visible, particularly in the top panels around the bright source Fornax\,A. In the bottom panels, enhanced noise is visible towards the bottom due to the effects of the fall-off in sensitivity towards the edge of the primary beam. 
The greyscale runs from 0 to 1\,\jypbm. The 90\%\ ANTARES error circles (radius 1\degr) are shown.
}
\end{figure*}

We extracted square image cutouts 5\degr\ on a side centered on the trigger positions, and combined these, taking the median value at each pixel position, to create a deep image for each trigger (center panels of Figure~\ref{fig:deep}). We also measured the RMS flux density in the object-subtracted background sky, \sigback\ (which corresponds to the sensitivity), in the $5\degr \times 5\degr$ regions centered on each trigger. The flux density for the faintest detectable source was set at 4\sigback.

The mean \sigback\ of the 33 \antth\ prompt snapshot images was 48\,\mjypbm, and the standard deviation of \sigback\ for these images was 4\,\mjypbm. The 34 \antfo\ prompt snapshots had $\sigback = 87 \pm 7$\,\mjypbm.  
For \antth, \sigback\ for the deep image made from the 33 snapshots should naively correspond to $48 / \sqrt{33} = 8$\,\mjypbm. After the snapshots have been median combined  (i.e.\ the median at each pixel is used), however, the measured \sigback\ is somewhat higher (18\,\mjypbm) than the naive expectation, due to the presence of sidelobes and confused sources \citep{gleam}.
Similarly, for \antfo\ we obtained 47\,\mjypbm\ for the deep median-combined image. The difference in sensitivity between the two fields is partly because of the difference in frequencies, and partly because \antfo\ is closer to the edge of the MWA primary beam than \antth, resulting in higher \sigback.

\subsection{Pre-trigger Comparison Images}\label{sec:pre}

We also retrieved archival MWA data from $\sim 20$ days prior to each trigger. For \antth\ we obtained 30 observations at 154\,MHz from UT 2013 Nov 1. For \antfo\ we obtained 31 observations  at 182\,MHz from UT 2014 Mar 2. These were analyzed in the same manner as described above. Deep images made from combining the $\sim 1$\,hr of observations for each trigger are shown in the left panels of Figure~\ref{fig:deep}. Comparison of the pre-trigger and prompt deep images by eye again showed no obvious transients. 

\begin{figure*}[htp]
\centering
\includegraphics[width=0.45\linewidth,draft=false]{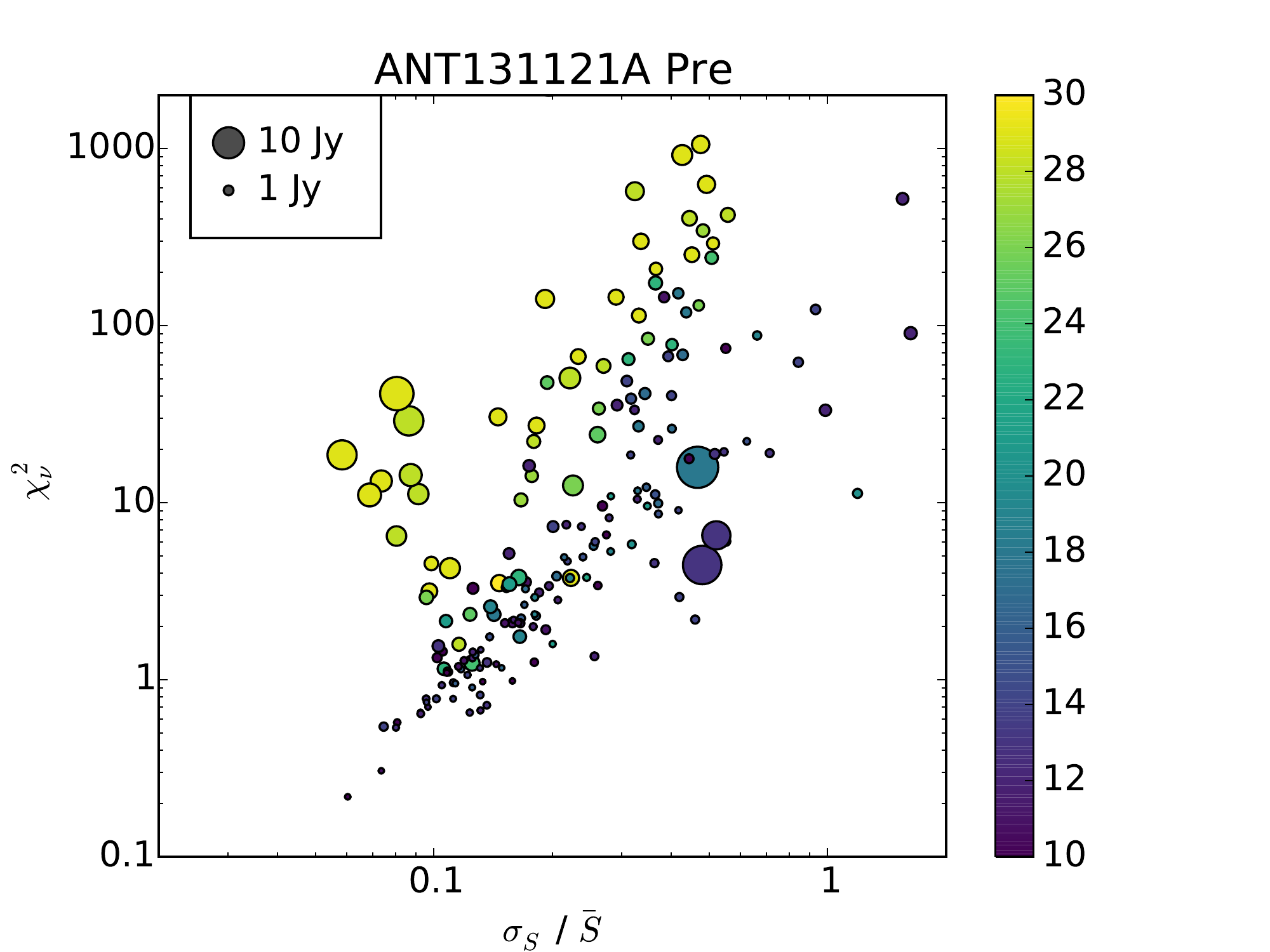}\quad%
\includegraphics[width=0.45\linewidth,draft=false]{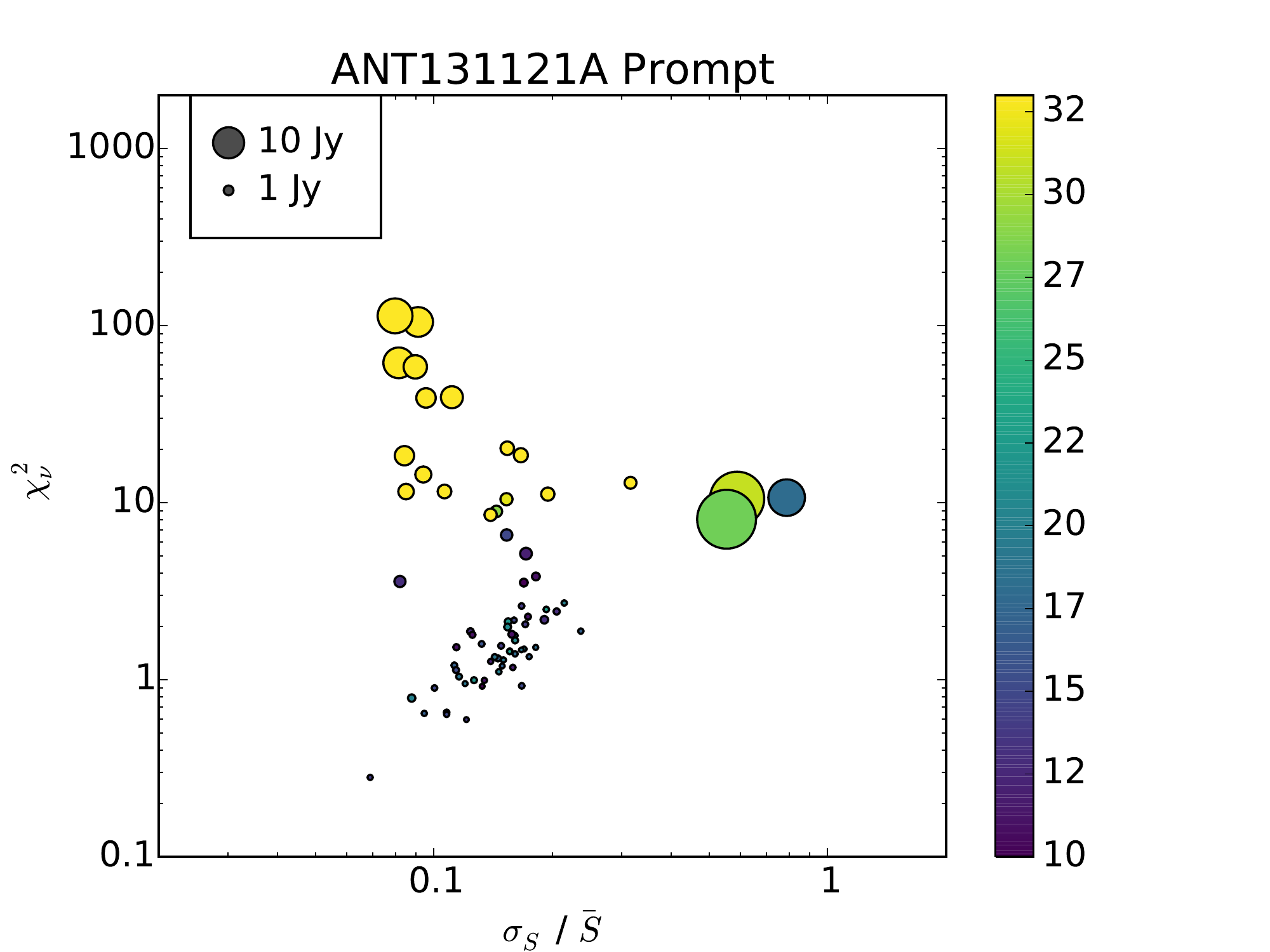}
\includegraphics[width=0.45\linewidth,draft=false]{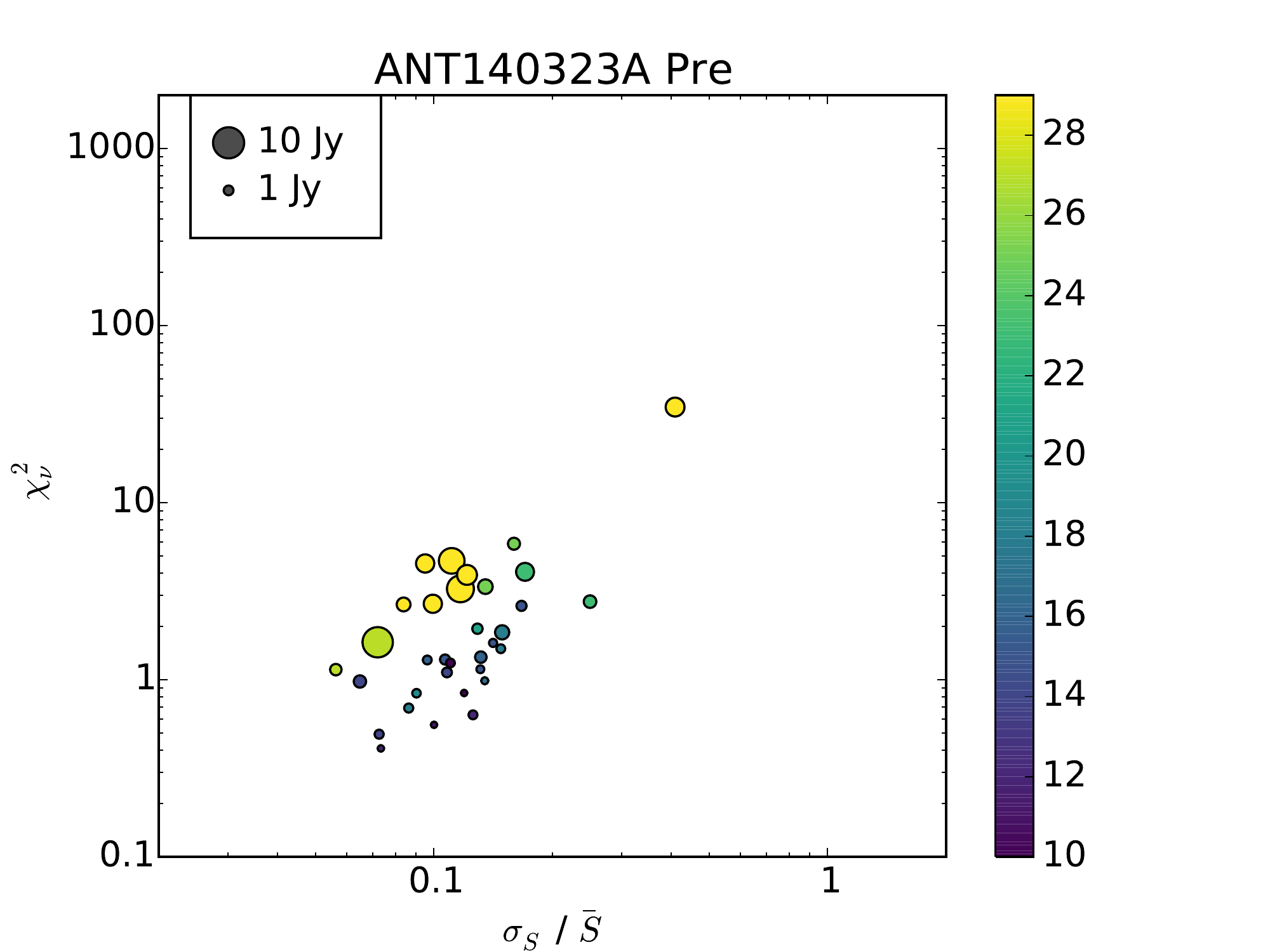}\quad%
\includegraphics[width=0.45\linewidth,draft=false]{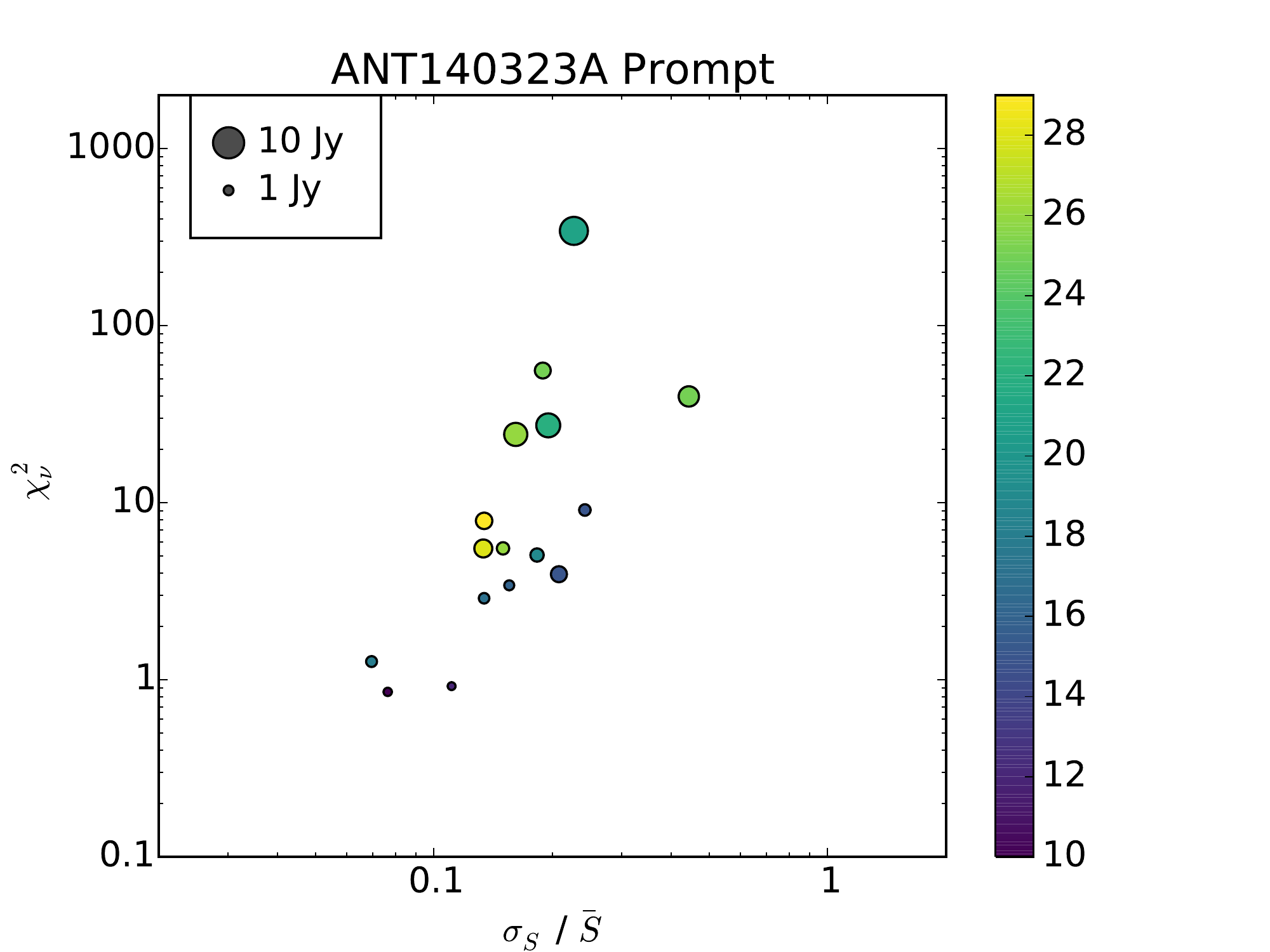}
\caption{\label{fig:var}
Reduced chi-squared, \rchisq, for the hypothesis that sources do not vary over the $\sim 30$ snapshots, plotted against fractional modulation (standard deviation, $\sigma_{\rm S}$, divided by the mean, $\bar{\rm S}$, flux density for the same sources). Sources are color-coded according to the number of snapshots in which they were detected. Data are plotted for pre-trigger (left) and prompt (right) datasets, for \antth\ (top) and \antfo\ (bottom). Only sources detected in $\geq 10$ of the snapshot images for each dataset are shown. Circle sizes scale with $\bar{\rm S}$.
}
\end{figure*}

We used the matched snapshot catalogs (independently for the pre-trigger and prompt datasets) to measure the mean ($\bar{\rm S}$) and standard deviation ($\sigma_{\rm S}$) of the flux densities of radio sources detected in our data. For all sources detected in at least 10 of the $\sim 30$ snapshots, we computed variability statistics (reduced chi-squared, $\rchisq$, and fractional modulation, $\fracmod$).  In Fig.~\ref{fig:var}, we plot \rchisq\ versus \fracmod\ for these sources. Varying image quality and detection thresholds make the comparison challenging, but if a trigger was associated with strong variability in an existing radio source, we might expect to see an outlier with high \rchisq\ and fractional modulation in the prompt dataset, but not in the corresponding pre-trigger dataset.

The majority of points in our variability plots occupy a contiguous region of parameter space, with brighter sources tending to be detected in more snapshots, and having higher \rchisq, as would be expected given improved signal to noise for these sources. Very few well-detected sources (those seen in $\sim 30$ snapshots) exhibit $\fracmod \gtrsim 50$\%, with the exception of the largest (i.e. brightest) two points in the \antth\ prompt plot, which have $\fracmod = 0.59$ and $0.55$ respectively. Both have $\rchisq \approx 10$, suggesting that they are indeed strongly variable. However, both are coincident with the lobes of Fornax\,A, and while variability of an AGN core is possible on short timescales, variability of extended lobe emission is not. We conclude therefore that the apparent variability here is caused by the difficulty of fitting point source models to extended emission. Variations in sensitivity and image quality result in different fits at each epoch, which is also why these sources do not appear in the same position in the top left panel of Figure~\ref{fig:var}. In any case, Fornax\,A is too far from the trigger position, given the ANTARES positional uncertainties, to be the neutrino source (likelihood of association $\sim 5 \times 10^{-4}$).

The plots for \antfo\ show fewer sources, due to the poorer sensitivity associated with the location of this candidate towards the edge of the primary beam. Nevertheless there are no well-detected sources that appear as outliers in the prompt data and not in the pre-trigger data. We conclude, therefore, that our observations did not convincingly detect any strong AGN flares associated with the neutrino triggers.

\subsection{Search for Late-Time Emission}\label{sec:post}

The MWA observing strategy, in particular changes in programs from one season to the next, somewhat restricts our ability to obtain a long-timescale follow-up of any given position of interest by simply searching the archive (as opposed to undertaking a dedicated follow-up campaign). Nevertheless, we were able to retrieve observations for both triggers that can be used to constrain late-time emission. We searched the archive for observations evenly distributed in log(time): 1, 2, 4, \ldots, 8192\,hr after the trigger. In most cases we were able to find data close in time to the desired epoch (Table~\ref{tab:rms}). When no suitable data were present in the archive that were closer in log(time) to a given epoch than to the previous or next epoch, that epoch was skipped.

Images were produced in the same manner as described above. Snapshot image sensitivity (which can be sensitive to the inclusion of relatively small amounts of poor quality data), \sigback, ranged from $49 - 373$\,\mjypbm\ (Table~\ref{tab:rms}). Once again, we made deep images (right panels of Figure~\ref{fig:deep}) by median-combining snapshots. Since the snapshots were taken over a wide range in time (see Table~\ref{tab:rms}) the median will de-emphasize sources which vary with a characteristic timescale $\ll 1$\,yr. These images nevertheless provide good sensitivity to long-timescale transient or variable sources associated with the neutrino.

Once again, neither trigger had an obvious transient counterpart, either in the snapshots, or the deep images. 

\begin{deluxetable}{llrrr}
\tablewidth{0pt}
\tabletypesize{\scriptsize}
\tablecaption{\label{tab:rms} Limits on radio counterparts}
\tablehead {
\colhead{UT date} &
\colhead{UT time} &
\colhead{Time since} &
\colhead{Frequency} & 
\colhead{\sigback}\\
\colhead{} &
\colhead{} &
\colhead{trigger (hr)} &
\colhead{(MHz)} & 
\colhead{(\mjypbm)}
}
\startdata
\cutinhead{\antth}
2013 Nov 1 average\tablenotemark{a} & 18:15:12 & -477 & 154 & 24 \\
2013 Nov 21 average\tablenotemark{a} & 14:50:56 & 0 & 154 & 18\\
2013 Nov 21\tablenotemark{b} & 16:59:04 & 2 & 154 & 49\\ 
2013 Nov 21\tablenotemark{b} & 18:36:40 & 4 & 154 & 93\\ 
2013 Nov 22\tablenotemark{b} & 14:47:04 & 24 & 154 & 51\\ 
2013 Nov 26\tablenotemark{b} & 18:16:56 & 123 & 154 & 201\\ 
2013 Dec 3\tablenotemark{b} & 14:18:56 & 287 & 154 & 50\\ 
2013 Dec 6\tablenotemark{b} & 15:25:28 & 360 & 154 & 200\\ 
2014 Feb 14\tablenotemark{b} & 13:00:56 & 2038 & 182 &  337\\ 
2014 Jul 20\tablenotemark{b} & 22:20:24 & 5791 & 182 & 48\\ 
2014 Oct 28\tablenotemark{b} & 16:48:24 & 8185 & 118 & 70\\ 
Late-time average\tablenotemark{a} & \ldots & 2 -- 8185 & 118, 154, 182 & 34\\
\cutinhead{\antfo}
2014 Mar 2 average\tablenotemark{a} & 16:20:16 & -503 & 182 & 36 \\
2014 Mar 23 average\tablenotemark{a} & 15:21:12 & 0 & 182 & 47\\
2014 Mar 24\tablenotemark{b} &12:36:40 & 21 & 154 & 133\\ 
2014 Mar 24\tablenotemark{b} &16:24:24 & 24 & 154 & 176\\ 
2014 Mar 26\tablenotemark{b} &12:28:48  & 68 & 154 & 143\\ 
2014 Mar 28\tablenotemark{b} &16:08:40 & 120 & 154 & 196\\ 
2014 Apr 3\tablenotemark{b} &11:57:20 & 260 & 154 & 135\\ 
2014 Apr 13\tablenotemark{b} &15:05:44 & 503 &154 & 178\\ 
2014 May 5\tablenotemark{b} &10:31:12 & 1027 & 154 &93\\ 
2014 Oct 25\tablenotemark{b} & 21:34:00 & 5190 & 154 & 373\\ 
2015 Feb 26\tablenotemark{b} &15:41:20 & 8160 & 154 & 68\\ 
Late-time average\tablenotemark{a} & \ldots & 21 -- 8160 & 154 & 68\\
\enddata
\tablenotetext{a}{Deep images made from median combining snapshot images.}
\tablenotetext{b}{Individual late-time snapshot images.}
\end{deluxetable}

\section{Limits on Progenitors}\label{sec:limits}

ANTARES detects $\sim 2$ atmospheric neutrinos per day with energies comparable to our two events ($\gtrsim 1$\,TeV). However, both candidates were generated by the ANTARES directional trigger (Section~\ref{sec:neutrino_events}), having positions coincident with galaxies within 20\,Mpc. Such coincidences represent $\sim 2$\% of the background from atmospheric events \citep{antaresearly}. If we assume the ANTARES neutrinos are indeed astrophysical, rather than due to terrestrial backgrounds, we can use our data to place some of the first low-frequency radio limits on EM counterparts to neutrino events. If the nearby galaxies are the hosts of the neutrino progenitors, this allows us to place limits on the luminosity of any EM counterpart.

Using $5\sigma$ upper limits of $90 - 340$\,mJy (based on \sigback\ for the deep images in Table~\ref{tab:rms}, which ranges from 18 to 68\,\mjypbm), we obtain $L_{150\,{\rm MHz}} \lesssim 10^{29}$\,\esh\ ($\lesssim 10^{37}$\,erg s$^{-1}$) for progenitors at 20\,Mpc. These limits are not strongly constraining of late-time emission from even the most luminous radio supernovae or GRBs at these distances; during the first $\sim 100$ days after the event, radio emission at MWA frequencies would be expected to be $\lesssim 10^{28}$\,\esh\ \citep{soderberg:00}. Our limits are better ($\lesssim 10^{27}$\,\esh) if \antfo\ is associated with the Antlia Dwarf at 1.3\,Mpc, but this still does not provide a strong constraint on progenitors. In fact, due to synchrotron self-absorption at low radio frequencies, late-time emission tends to be faint in general \citep[e.g.][]{metzger:15}, further emphasizing the need for rapid response or simultaneous observations to search for brighter prompt radio emission.

For GRBs or CCSNe at distances $< 20$\,Mpc, we consider whether counterparts should have been seen in the optical observations (Section~\ref{sec:neutrino_events}). At 20\,Mpc, the optical limit of 18.7\,mag corresponds to absolute magnitudes brighter than $-13$, sensitive enough to detect all but the faintest \citep[e.g.][]{pastorello:07} supernovae, although this does not account for dust obscuration in the host galaxy. We also consider a scenario where the nearby galaxies are chance alignments, and the progenitors are in fact at larger distances. Considering the possibility that the neutrinos might be from binary neutron star coalescences such as those modeled by \citet{pshirkov}, our upper limits for prompt emission, with their Equation~8 and assuming an efficiency scaling exponent $\gamma = 0$, would place such progenitors at distances $\gtrsim 1$\,Gpc ($z \gtrsim 0.2$). 

\section{Outlook}\label{sec:future}

 Although the MWA has excellent capabilities for these kinds of serendipitous searches due to its wide field of view, the use of archival data has limitations. Neither trigger was optimally placed within the MWA field of view: \antth\ was $\sim 8$\degr\ from the pointing center, and \antfo\ was $\sim 17$\degr\ away. Particularly in the latter case, the fall-off in primary beam response means that noise in the region of the image near the trigger position is higher than is ideal. Going forward, we intend to trigger pointed observations soon after a neutrino detection. The region of sky seen from the MWA is well-matched to where ANTARES has good sensitivity, meaning that around 40\%\ of ANTARES upward-going events are accessible to rapid MWA follow-up. ANTARES can generate triggers in a few seconds, and MWA can point at the trigger position within another 10\,s, allowing us to probe dispersion measures as low as 100\,pc\,cm$^{-3}$ \citep{grb150424a}, sufficiently fast to detect even minimally-dispersed events from the nearest galaxies. The MWA's wide field of view also means that targeted follow-up observations easily probe the entire error circle of ANTARES events with optimal MWA sensitivity.   

It is notable that Fornax\,A, one of the brightest radio sources in the sky (associated with NGC\,1316 at a distance of $\sim 20$\,Mpc) is close ($\sim 3$\degr) to the position of \antth, although it is strongly ruled out as the progenitor given the positional uncertainties of the ANTARES trigger. However, this region of the sky is densely populated with galaxies (including $\sim 10$ bright members of the Fornax Cluster within the neutrino error circle), illustrating the importance of EM observations coincident in time with neutrino triggers to resolve ambiguity as to the progenitor. 

Although we found no strongly varying radio counterpart to the two triggers discussed here, MWA data at the positions of additional ANTARES triggers exist in our archive, albeit not simultaneous in time with the triggers. We defer the analysis of late-time and pre-trigger observations of these events to a future paper. Additionally, future rapid follow-up (a capability already demonstrated at MWA), combined with an increase in sensitivity (due to a decrease in the confusion limit from the recently approved MWA expansion), mean that MWA is well placed to follow up promising neutrino candidates over the next few years.

\acknowledgments

This scientific work makes use of the Murchison Radio-astronomy Observatory, operated by CSIRO. We acknowledge the Wajarri Yamatji people as the traditional owners of the Observatory site. Support for the operation of the MWA is provided by the Australian Government Department of Industry and Science and Department of Education (National Collaborative Research Infrastructure Strategy: NCRIS), under a contract to Curtin University administered by Astronomy Australia Limited. We acknowledge the iVEC Petabyte Data Store and the Initiative in Innovative Computing and the CUDA Center for Excellence sponsored by NVIDIA at Harvard University. DLK and SDC acknowledge support from the US National Science Foundation (grant AST-1412421).
We acknowledge the financial support of the funding agencies: Centre National de la Recherche Scientifique (CNRS), Commissariat \`a l'\'ene\-gie atomique et aux \'energies alternatives (CEA), Commission Europ\'eenne (FEDER fund and Marie Curie Program), R\'egion Ile-de-France (DIM-ACAV), R\'egion Alsace (contrat CPER), R\'egion Provence-Alpes-C\^ote d'Azur, D\'epartement du Var and Ville de La Seyne-sur-Mer, France; Bundesministerium f\"ur Bildung und Forschung (BMBF), Germany; Istituto Nazionale di Fisica Nucleare (INFN), Italy; Stichting voor Fundamenteel Onderzoek der Materie (FOM), Nederlandse organisatie voor Wetenschappelijk Onderzoek (NWO), the Netherlands; Council of the President of the Russian Federation for young scientists and leading scientific schools supporting grants, Russia; National Authority for Scientific Research (ANCS), Romania; Ministerio de Ciencia e Innovaci\'on (MICINN), Prometeo of Generalitat Valenciana and MultiDark, Spain; Agence de l'Oriental and CNRST, Morocco. We also acknowledge the technical support of Ifremer, AIM and Foselev Marine for the sea operation and the CC-IN2P3 for the computing facilities. We gratefully acknowledge financial support from the OCEVU LabEx, France.
We thank Nathan Whitehorn 
and the anonymous referee
for helpful comments.

 \bibliographystyle{apj}

 \end{document}